\documentclass[aps,preprint]{revtex4-1} 
\usepackage{graphicx}
\usepackage{graphics}
\usepackage{amsmath}

\newcommand{\half}{\frac{1}{2}}

\newcommand{\eps}{\epsilon}

\newcommand\refeq[1]{(\ref{#1})}
\newcommand\reffig[1]{Figure~\ref{#1}}
\newcommand\of[1]{\left( #1 \right)}
\newcommand\sqof[1]{\left[ #1 \right]}
\newcommand\vecbf[1]{{\bf #1}}

\begin{document}

\title{Charged Radial Infall for Spherical Central Bodies}

\author{J. Franklin}
\email{jfrankli@reed.edu}
\affiliation{Department of Physics, Reed College, Portland, Oregon 97202,
USA}
\author{F. Morton-Park}
\affiliation{Department of Physics, Reed College, Portland, Oregon 97202,
USA}

\begin{abstract}
A massive, charged, spherical central body can be neutralized by attracting particles of opposite charge.  We calculate the time it takes to neutralize these central bodies using classical mechanics, special relativistic mechanics, and finally, the ``forced" trajectories of general relativity.  While we can compare the classical and (special) relativistic times, and find, predictably, that the special relativistic neutralization time is longer, a comparison of these times with the general relativistic result is not as directly possible.  We offer the final calculation as a demonstration of dynamics in a general setting and in particular, the structural similarity of the (general) relativistic problem to the other cases.  
\end{abstract}

\maketitle

\section{Introduction}

The common lore, that one ``does not expect any macroscopic body to possess a net charge"~\cite{Chandra} is certainly compelling observationally.  But it invites the question, especially relevant during black hole formation:  ``How long does it take for a charged spherical body to neutralize?"  Here we answer that question in three different ways.   We consider the problem of charge neutralization for ``classical" Newton-Coulomb black holes, and compare the time required to neutralize a central spherical body to the time required for a (special) relativistic neutralization, and finally, a full general relativistic calculation.

In all cases, we have a specific model in mind -- take a central body of mass $M$ carrying charge $Q = N \, q$ where $q$ is some fundamental unit of charge (the charge of the constituent particles making up the central body, say) and $N$ is a large number.  The neutralization process consists of arranging $N$ test charges, each carrying charge $-q$, in a sphere of radius $R$ and allowing them to fall radially inward under the influence of both gravity and the Coulomb force.  Note that we ignore the force on the charges associated with the shell itself (which, as we discuss at the end, changes the ratio of Coulomb-to-gravitational interaction by a factor of one half).  We will take our model central body to be a black hole, so there is a natural radius $r_0$ associated with the end of the infall process, the Schwarzschild radius (in the un-charged case).  This model ignores the interaction of the ``test masses" forming the charged spherical shell, and since each test particle falls radially in, they all take the same time to reach the radius of the spherical body.  So we have a minimum time-to-neutralization, with no angular momentum, and no inter-particle interaction.  These simplifying assumptions mean that we can focus on a single test particle, and our work amounts to repeatedly finding the time it takes that particle to go from $R$ (at rest) to $r_0$ in various settings of increasing physical sophistication.

\section{Newton-Coulomb Black Holes}\label{NCBH}

Consider a sphere of radius $r$ with mass $M$ distributed uniformly throughout its volume.   The sphere sets up a gravitational force on massive particles -- we can define the escape speed for test particles as the minimum speed needed to start at the surface of the sphere and end at spatial infinity (at rest).  That is just the usual:
\begin{equation}\label{escape}
\half \, m \, v^2 = \frac{G \, M \, m}{r}.
\end{equation}
But this equation can, in a sense, be turned around -- instead of the escape speed $v$, we can ask for what radius $r_0$ does light itself ``just" escape the gravitational field of the central body?  From~\refeq{escape}, with $v = c$, we have:
\begin{equation}\label{radius}
r_0 = \frac{2 \, G \, M}{c^2}.
\end{equation}
This classical result is well-known: John Michell, the British geologist discussed this special radius in 1783, and Laplace independently developed the same idea in 1796, it is given in the first edition of {\it Exposition du Systeme du Monde} (although he removed it from subsequent editions).  More of the history of these early ``dark stars" can be found in~\cite{Israel}.

The applicability of the argument that takes us from~\refeq{escape} to~\refeq{radius} relies on canceling the $m$s appearing on either side of~\refeq{escape} prior to solving for $r_0$ -- otherwise, the massless photon returns $0 = 0$ trivially.  It is interesting, then, that the same fundamental length (associated with a given mass $M$) appears in general relativity, where $r_0$ is the ``event horizon" associated with the Schwarzschild spacetime.  The physical interpretation is also the same -- that is the radius at which not even light can escape the gravitational field of the central body. Both because of the classical argument, and its general relativistic relevance, we will take our classical central bodies to have radius $r_0 = \frac{2 \, G \, M}{c^2}$, and that will define the impact radius for the test particles that we send in to bring about the neutralization.

\begin{figure}[htbp] 
   \centering
   \includegraphics[width=3in]{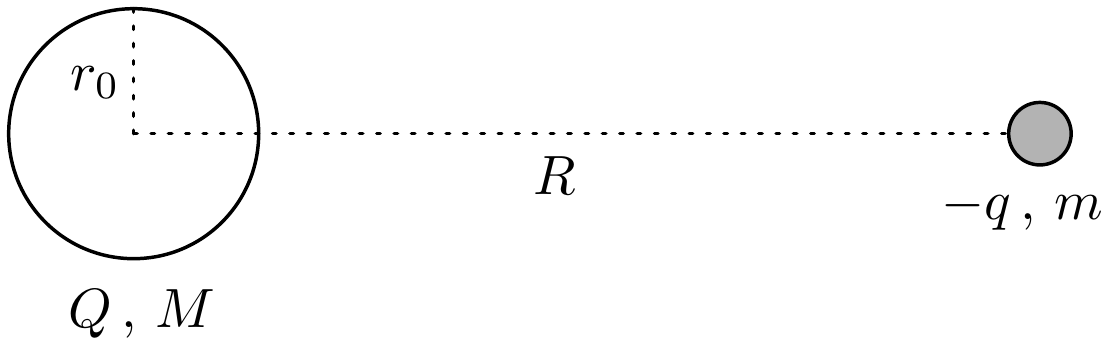} 
   \caption{A test mass $m$ carrying charge $-q$ is released from rest a distance $R$ from a central sphere of mass $M$ carrying total charge $Q$.  How long does it take for the test mass to reach the surface of the central sphere, $r_0$?}
   \label{fig:Setup}
\end{figure}

The problem we need to solve, then, is:  Given a uniformly charged sphere of radius $r_0$ carrying total charge $Q = N \, q$ and mass $M$, how long does it take a test particle of mass $m$ and charge $-q$ to reach $r_0$ if it starts from rest at $R$?  The setup is shown in~\reffig{fig:Setup}.

\subsection{Newtonian Result}

To answer this question in the classical setting, we could work directly from Newton's second law:
\begin{equation}
m \, \ddot r = -\frac{G \, M \, m}{r^2} - \frac{Q \, q}{4 \, \pi \, \eps_0 \, r^2}
\end{equation}
and integrate twice to find $r(t)$, then invert $r(t^*) = r_0$ to find the time $t^*$ at which the test particle collides with the central body.  It is slightly easier (and more generalizable for future calculations) to start from the Hamiltonian, representing the total energy of the system:
\begin{equation}\label{totalE}
E = \half \, m \, \dot r^2 - \frac{G\, M \, m}{r} - \frac{Q \, q}{4 \, \pi \, \eps_0 \, r}.
\end{equation}
Because we want to compare the charged case with the un-charged case, we define the force ratio $\beta \equiv \frac{Q \, q/(4 \, \pi \, \eps_0)}{G \, M \, m}$, so that $\beta = 0$ recovers the Newton (with no Coulomb) result.  Then~\refeq{totalE} can be solved for $\dot r^2$:
\begin{equation}
\dot r^2 = \frac{2 \, E}{m} + 2 \, G \, M \, \of{1 + \beta} \, \frac{1}{r}.
\end{equation}
Finally, we can render each term dimensionless by setting $r = \rho \, r_0$ and $t = \sigma \, t_0$ for dimensionless $\rho$ and $\sigma$.  Then 
\begin{equation}\label{rho2}
\rho'^2 = \frac{2 \, E}{m \, \of{\frac{r_0}{t_0}}^2} + \frac{2 \, G \, M}{r_0 \, \of{\frac{r_0}{t_0}}^2} \, \of{1 + \beta} \, \frac{1}{\rho}
\end{equation}
where $\rho' \equiv \frac{d \rho(\sigma)}{d \sigma}$.  Now $r_0$ is the Schwarzschild radius, but $t_0$ is some characteristic time that is up to us -- motivated by the first term on the right in~\refeq{rho2} and the definition $r_0 \equiv \frac{2 \, G \, M}{c^2}$, we set $\frac{r_0}{t_0} \equiv c$, and then:
\begin{equation}\label{rho2f}
\rho'^2 = \frac{2 \, E}{m \, c^2} + \of{1 + \beta} \, \frac{1}{\rho}.
\end{equation}

All that's left to do is encode the initial conditions:  Let $R = \eta \, r_0$, then $\rho(0) = \eta$, and $\rho'(0) = 0$.  Conservation of energy demands that:
\begin{equation}
0 = \frac{2 \, E}{m \, c^2} + \of{1 + \beta} \, \frac{1}{\eta} \longrightarrow E = -\half \, m \, c^2 \, \of{1+\beta} \, \frac{1}{\eta},
\end{equation}
so that~\refeq{rho2f} becomes
\begin{equation}\label{rho2ff}
\rho'^2 = \of{1 + \beta} \, \sqof{\frac{1}{\rho} - \frac{1}{\eta}}.
\end{equation}
The goal is to solve~\refeq{rho2ff} for $\rho(\sigma)$, and then find $\sigma^*$ such that $\rho(\sigma^*) = 1$, corresponding to $r(t^*) = r_0$.
If we rewrite~\refeq{rho2ff} in terms of $d\rho$ and $d\sigma$, 
\begin{equation}
d\sigma = -\frac{d\rho}{\sqrt{\of{1 + \beta}\, \of{\frac{1}{\rho} - \frac{1}{\eta}}}},
\end{equation}
we can integrate directly from $\sigma = 0$ to $\sigma^*$ on the left, $\rho = \eta$ to $1$ on the right, giving an expression for $\sigma^*$:
\begin{equation}\label{exacter}
\sigma^* = \frac{\sqrt{\eta} \, \sqof{ 4 \, \sqrt{\eta -1} + \eta \, \pi + 2 \, \eta \, \tan^{-1}\of{\frac{\eta-2}{2 \, \sqrt{\eta-1}}} } }
{4 \, \sqrt{1 + \beta}}.
\end{equation}

\begin{figure}[htbp] 
   \centering
   \includegraphics[width=3.5in]{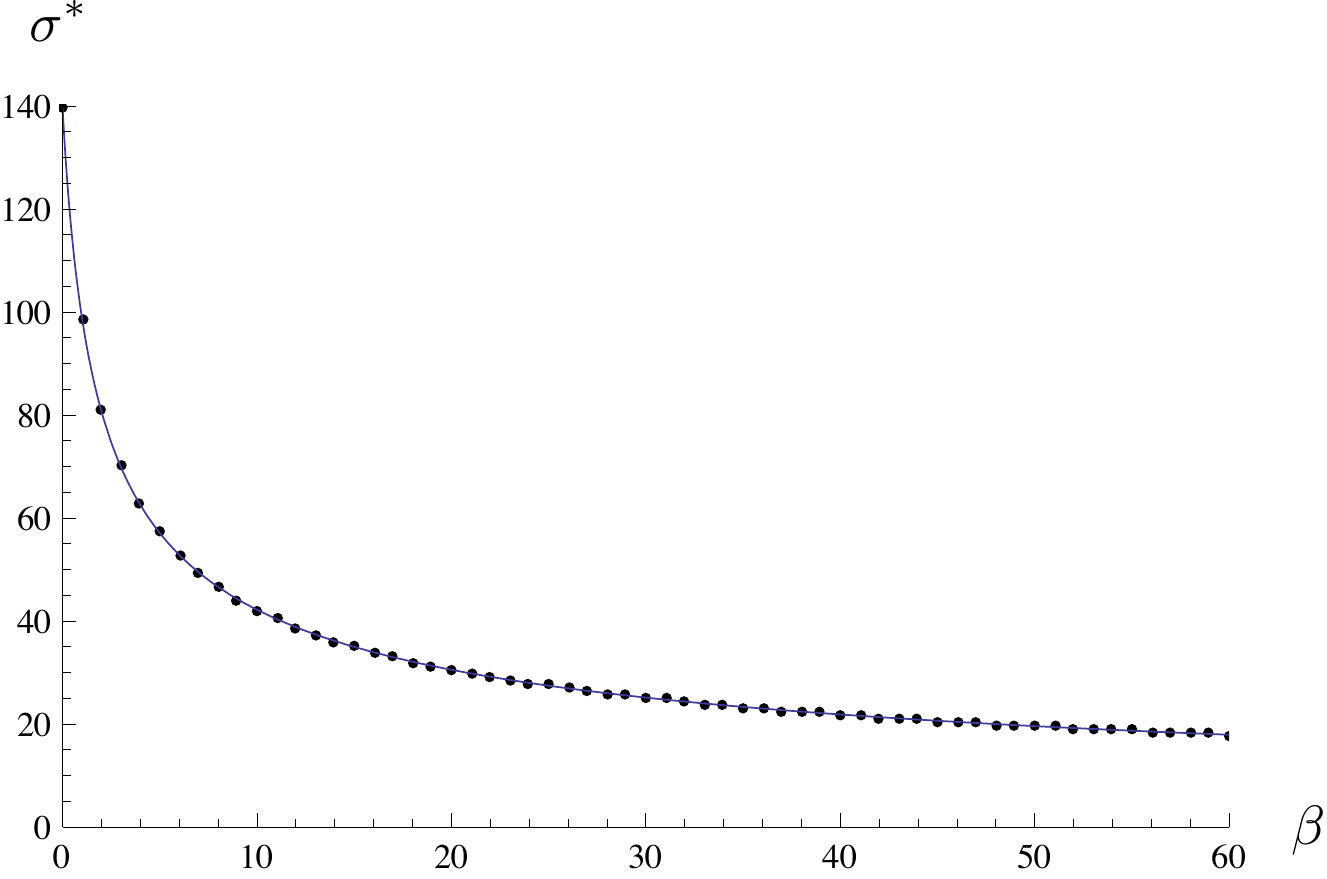} 
   \caption{Time to neutralization for a classical Newton-Coulomb central body with a test particle starting at $\eta = 20$.  The black points are the numerically computed time, the blue curve is the plot of~\refeq{exacter}.}
   \label{fig:NC}
\end{figure}

With this expression, we can compute the time it will take for a charged sphere to neutralize.   However, the particular solution found here is not extensible to the general relativistic version of the neutralization problem, and we would like to use the same approach in both the classical and relativistic cases.  From that point of view, we want a method that will allow us to determine $\sigma^*$ absent an explicit expression (ugly or not).  We use a Runge-Kutta routine with a root-finding procedure to solve~\refeq{rho2ff} for $\sigma^*$ such that $\rho(\sigma^*) = 1$ as a function of $\beta$~\cite{note}.  The result for $\eta = 20$ (so we start at a distance $R = 20 \, r_0$ from the center of the central body) is shown in~\reffig{fig:NC}.  As is to be expected, the neutralization time decreases as the electrostatic-to-gravitational force ratio gets larger (we are increasing the total magnitude of the force on the test particles).  For comparison and as a test of the method, the exact result~\refeq{exacter} is plotted in~\reffig{fig:NC} as well.  
 
As this is a classical calculation, the speed limit of special relativity will surely be violated at some point.  In these units, the speed of light is $1$, and it is clear from~\refeq{rho2ff} that when:
\begin{equation}\label{rhocut}
\of{1 + \beta} \, \of{\frac{1}{\rho} - \frac{1}{\eta}} > 1 \longrightarrow \rho < \frac{\eta\, \of{1 + \beta}}{\eta + \of{1 + \beta}}
\end{equation}
the speed of the test particle will be greater than $c$.   For $\beta = 0$, we see that $\rho' < 1$ for all values of $\rho$, since $\rho > 1$ (when $\rho = 1$, we are at $r = r_0$, and the test particle has collided with the central body) and $\frac{\eta}{\eta + 1} < 1$.  That makes sense when you think about how the radius $r_0$ was chosen, but if $\beta > 0$, it is possible that the inequality in~\refeq{rhocut} is achieved by values of $\rho > 1$, leading to particle speeds greater than that of light prior to collision with the central body.  In order to address this issue, we re-run the analysis starting from the total energy associated with special relativity.
\subsection{Special Relativity}

\begin{figure}[htbp] 
   \centering
   \includegraphics[width=3.5in]{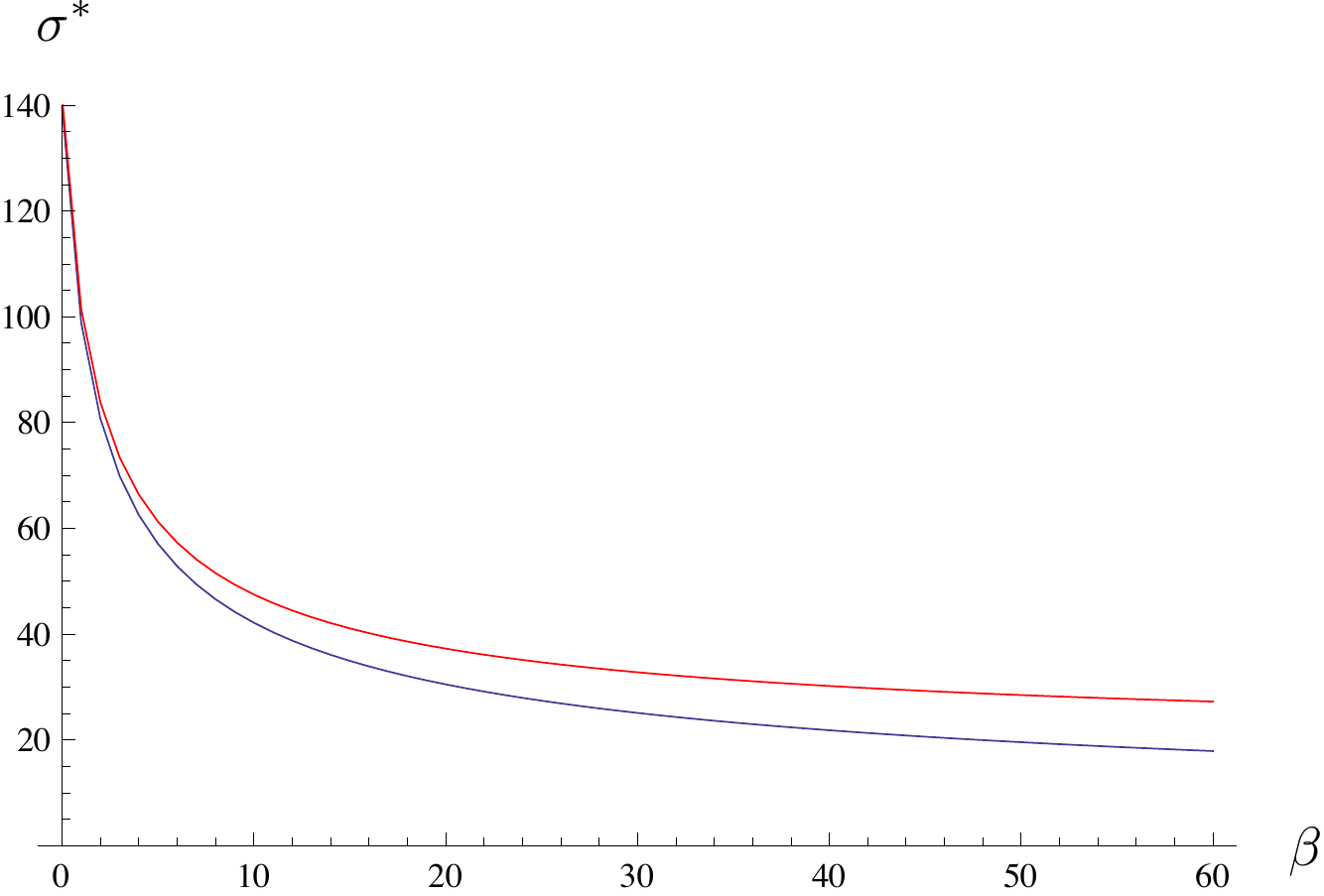} 
   \caption{In red, the relativistic time-to-neutralization for our model problem with $\eta = 20$, and in blue, the original non-relativistic result from~\reffig{fig:NC}.  At some point, the relativistic speed saturates, a feature not shared by the classical trajectories -- since the special relativistic particles must travel $<c$, they have an overall longer neutralization time.}
   \label{fig:NCSR}
\end{figure}

The equation replacing~\refeq{totalE} in the relativistic setting is:
\begin{equation}\label{specialE}
E = \frac{m \, c^2}{\sqrt{1 - \frac{\dot r^2}{c^2}}} - \frac{G \, M \, m}{r} - \frac{Q \, q}{4 \, \pi \, \eps_0 r}
\end{equation}
where the rest energy of the test particle is now included in $E$ (and the rest of relativistic dynamics is built in).  In the dimensionless variables of the previous section, this equation reads:
\begin{equation}
\frac{E}{m \, c^2} = \frac{1}{\sqrt{1 - \rho'^2}} -\half \,  \of{1 + \beta} \, \frac{1}{\rho}
\end{equation}
and we can solve for $\rho'^2$ to construct the analogue of~\refeq{rho2f},
\begin{equation}
\rho'^2 = 1 - \frac{1}{\sqof{ \frac{E}{m \, c^2} + \half \, \of{1 + \beta} \, \frac{1}{\rho}}^2}.
\end{equation}
Once again imposing the initial condition:
\begin{equation}
0 = 1 -  \frac{1}{\sqof{ \frac{E}{m \, c^2} +\half \, \of{1 + \beta} \, \frac{1}{\eta}}^2} \longrightarrow E = m \, c^2 \, \of{1- \half \, \of{1 + \beta} \, \frac{1}{\eta} }
\end{equation}
so that
\begin{equation}
\rho'^2 = 1 - \frac{1}{\sqof{1 + \half \, \of{1+\beta} \, \of{\frac{1}{\rho} - \frac{1}{\eta}}}^2}.
\end{equation}
Notice that this equation enforces a speed limit of $1$ as it should~\cite{DGnote}.  The neutralization times for the model problem are shown in red in~\reffig{fig:NCSR} -- as the force ratio increases, the special relativistic version has neutralization times that are {\it longer} than the classical calculation.  This makes sense, since classically, particles can travel faster than $c$, while the relativistic speed limit \ldots limits particle speeds.  When the attractive force is greater, a larger portion of the classical trajectory has particle speed greater than $c$, while the relativistic trajectory always has speed $< c$, leading to a longer time.

There is, of course, a natural limit in the case of special relativity -- the test particles need to travel a distance $R - r_0$ and the maximum speed at which they could do so would be $c$, leading to 
\begin{equation}
t^* = \frac{R - r_0}{c}
\end{equation}
or, in our current units, $\sigma^* = \eta - 1$.

 The Newtonian potential used in~\refeq{specialE} is a natural starting point for relativistic investigation, but the special relativistic case presented here does not take into account the fact that Newtonian gravity, unlike E\&M, is not consistent with special relativity.  The point-source solution we have used solves Laplace's equation, and therefore implies that the gravitational force changes instantaneously with a change in the source properties.  While not an immediate concern for this static configuration, it is a hallmark of bad behavior under Lorentz transformation (in E\&M, the magnetic field saves the special relativistic character of Maxwell's equations, and there is an analagous approximate force that can be introduced to perturbatively correct the Newtonian gravitational field~\cite{Baker}).  Indeed, the reconciliation of gravity with special relativity leads to general relativity.

\section{Relativistic Black Holes}

In general relativity (GR), Einstein's equations relate the field of the theory to sources just as Maxwell's equations do for E\&M.  There are deep structural differences between the two, but the story we tell ourselves: ``Sources generate fields" is the same.  The field of GR (the target functions, like the potentials $V$ and $\vecbf A$ of E\&M) is the metric of spacetime, and the sources are mass and moving mass (more generally, since mass and energy are equivalent, any form of energy or moving energy).  Once Einstein's equations have been solved, the metric influences the motion of test particles, just as once you have solved for $V$ and $\vecbf A$, you can find the force on test particles from the Lorentz force law in E\&M.  The nature of that influence, for gravity, is different, but the end result is a Hamiltonian similar to~\refeq{specialE}, depending in a more complicated way on the location of the test particle.  The total energy replacing~\refeq{totalE} and~\refeq{specialE} in general relativity, for a massive central body carrying charge $Q$ (the so-called 
Reissner-Nordstrom \cite{RN} solution to Einstein's equation) is:
\begin{equation}
\begin{aligned}
E &=  m \, c^2 \, \frac{1 - \frac{r_0}{r} + \frac{B}{r^2}}{\sqrt{1 - \frac{r_0}{r} + \frac{B}{r^2} - \frac{\dot r^2/c^2}{1 - \frac{r_0}{r} + \frac{B}{r^2}} }} - \frac{q \, Q}{4 \, \pi \, \eps_0 \, r} \\
B &\equiv \frac{G \, Q^2}{4 \, \pi \, \eps_0 \, c^4}
\end{aligned}
\end{equation}
As a check, you should try expanding the above in $r$ and verify that you recover~\refeq{specialE}.  Notice that we only introduced the Coulomb potential as ``additional" energy -- the first term in the Hamiltonian is associated with the rest energy, the kinetic energy, and a ``new" relativistic energy (reducing to the Newtonian gravitational potential in the appropriate limit).

\begin{figure}[htbp] 
   \centering
   \includegraphics[width=3.5in]{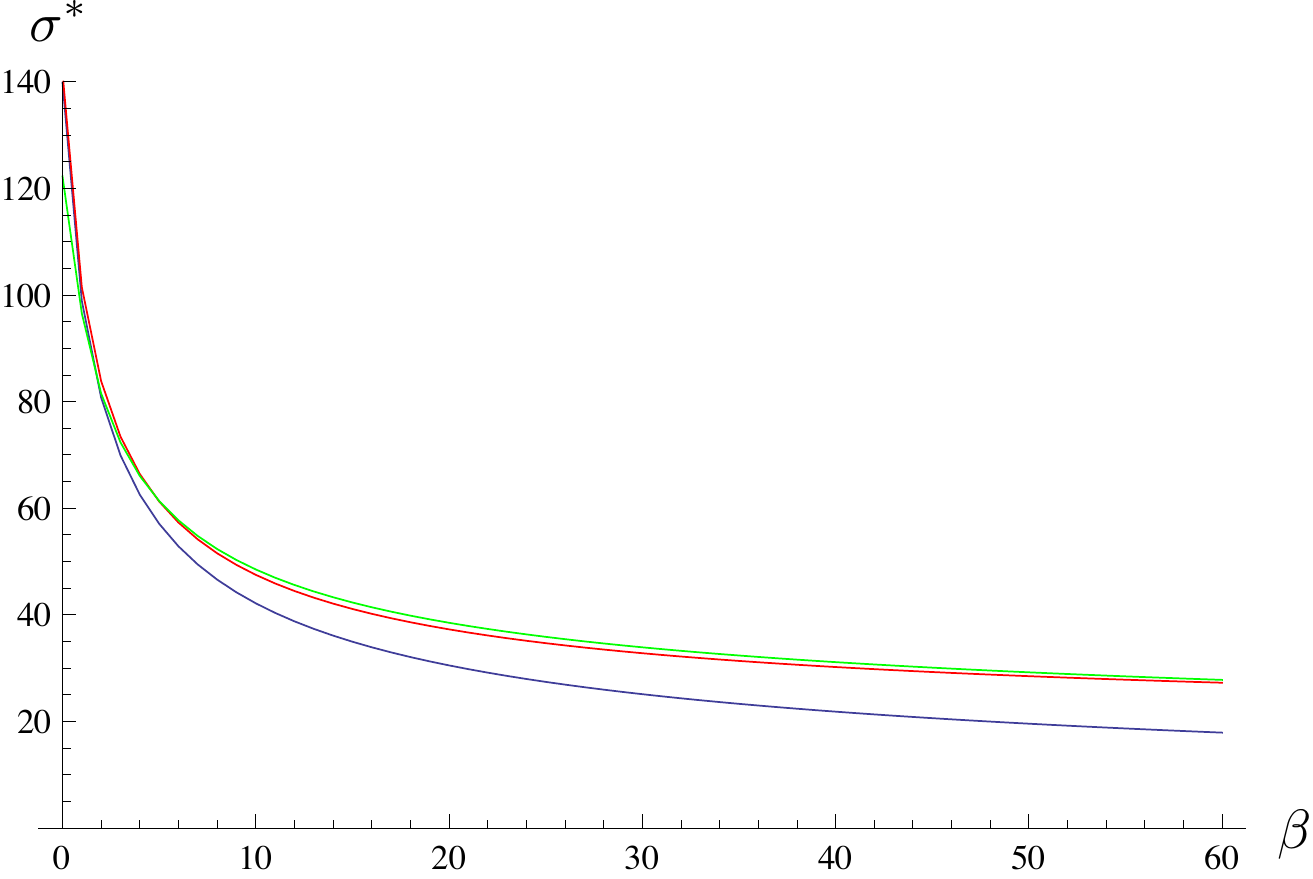} 
   \caption{In green, the time-to-neutralization for the Reissner-Nordstrom relativistic solution.  In red and blue are our previous special relativistic and classical results for comparison.}
   \label{fig:GRTTN}
\end{figure}

Now using our usual dimensionless variables, we have:
\begin{equation}
\begin{aligned}
E &= m \, c^2 \, \frac{1 - \frac{1}{\rho} + \frac{\bar\beta}{4 \, \rho^2}}{\sqof{1 - \frac{1}{\rho} + \frac{\bar \beta}{4 \, \rho^2} - \frac{\rho'^2}{1 - \frac{1}{\rho} + \frac{\bar \beta}{4 \, \rho^2}}}^{1/2}} - m \, c^2 \, \frac{\beta}{2 \, \rho}  \\
\bar\beta &\equiv \frac{ \frac{Q^2}{4 \, \pi \, \eps_0}}{G \, M^2}.
\end{aligned}
\end{equation}
The new parameter $\bar\beta$ is again a ratio of electrostatic to gravitational forcing, this time with the central body's parameters only.  We know that $Q = N \, q$, and if we take $M = N\, m$ (i.e. the test particles are precisely the charged particles making up the central body), then we have $\bar \beta = \beta$, and we can proceed with our usual routine.  The resulting time-to-neutralization, as a function of $\beta$ is shown in green in~\reffig{fig:GRTTN}. 

In order to generate a roughly comparable time-to-neutralization here, we must be careful to define the starting point.  In general relativity, the distance between two points (length) is determined by 
the metric of spacetime, and this metric has different values based on our choice of $\beta$.  If we insist that the (radial) distance travelled by the test particles be $\eta_i - \eta_f = 19$ ($\eta_f = 1$ by definition of our problem, and we start the particles at $\eta_i = 20$), then each value of $\beta$ used to generate~\reffig{fig:GRTTN} requires a different starting point.  In particular, we want $\ell = 19$, for:
\begin{equation}\label{ellsame}
\ell = \int_1^{\eta^*} \sqof{1 - \frac{1}{\rho} + \frac{\beta}{4 \, \rho^2} }^{-1/2} \, d\rho,
\end{equation}
where $\ell$ is the proper length associated with zero temporal separation in the Reissner-Nordstrom spacetime (i.e. a length calculated instantaneously).  We can use this equation to determine the label of our starting point, $\eta^*$, in the modified spacetime setting~\cite{GRnote}.

\section{Conclusion}

We have presented a simple model for the neutralization of a charged central body and carried out both the classical and special relativistic analysis for a variety of charge-to-mass ratios.  The time it takes to neutralize a central body using special relativistic dynamics is longer than the corresponding time calculated from Newton's second law since in the relativistic case, particles cannot travel faster than $c$.  In addition, we set up and solved the analogous neutralization problem in general relativity, using the Reissner-Nordstrom metric to generate equations of motion for a charged test particle in that setting.    It is interesting that while the general relativity form of the problem involves a variety of new physical ideas, the actual execution of the calculation is remarkably similar to its classical and special relativistic counterparts.  As long as our test particles start far away from the central body, the general relativistic limit will match the special relativistic one (the test particles stay away from the strong field region of the spacetime, where interpretation is not straightforward).  In our case, starting $20$ Schwarzschild radii away led to virtually identical results for special and general relativity.

Our model for neutralization was the simplest available -- more sophisticated approaches, like that taken in~\cite{FMP} where the change in charge of the central body as the neutralization occurs is considered, provide additional physical insight when the classical and relativistic cases are compared.   In addition, we could consider the force associated with the neutralizing sphere itself -- that changes what we mean by $\beta$ in a particular model by introducing the repulsive shell force: $\half \, \frac{N \, q^2}{4 \, \pi \, \eps_0 \, r^2}$ for each particle \cite{Griffiths}.  This exercise, on the classical side, would provide a good introduction to the problem of spherical collapse in general relativity, where that same ``self-force" is interesting \cite{ShellOne, ShellTwo}.  Indeed in the general relativistic case, collapse depends on the details of the dust elements falling into the central black hole -- the self-gravity of these cannot be ignored, and is an integral part of the study of astrophysically relevant collapse.   Even our modest toy problem becomes more involved in the full GR case.

In addition to the above simplifications, where we ignore the interaction of the test particles with each other, we have also left out the interaction of the test particles with themselves.  There are electromagnetic radiative corrections that will tend to slow the particle's progress, as energy is lost to the radiation fields.  These corrections can be calculated in the classical and relativistic cases using familiar electrodynamics notions  (as in Chapter 11 of~\cite{Griffiths}).

In our simplified setting, we can put some relatively concrete bounds on the neutralization time.  Taking the special relativistic result, in the large $\beta$ limit, we know that the minimum possible time is $\sigma^* = \eta - 1$, or, with units:
\begin{equation}
t^* = \of{\eta - 1}\, \frac{r_0}{c}.
\end{equation}
where $\eta$ is the number of Schwarschild radii away we use as the starting point, and $r_0$ is the Schwarzschild radius of a neutral black hole~\cite{GRnote2}.  The minimum time scales like $\eta$ -- we expect $r_0$ to be a small number -- if we take a supermassive black hole with a mass of, say, $10^6 \, M_\circ$, then $r_0 \sim 10^6$ km and the time it takes for light to travel $20 \, r_0$ is about a minute, instantaneous on the timescale of the universe.  Of course, the extreme special relativistic limit supposed that $\beta$ is large, we can imagine much longer times, up to the amount of time it takes for uncharged material to fall radially into the black hole (the $\beta = 0$ limit).

From our assumption that $\beta = \bar\beta$, in which the infalling matter is the same as that constituting the central body, we have restricted ourselves, in the general relativistic version of the problem.  For the Reisnner-Nordstrom spacetime, we must have $Q \le M$ (in units where $G = c = 1$), so the maximum value of $Q$ is $M$, giving $\beta \le 1$, the neutralization is then dominated by the charge-free case.  If we relax this requirement, then we can have $\bar\beta = 1$, say (for a maximally charged central body), and we are still free to choose $\beta = \frac{q}{\sqrt{4 \, \pi \, \eps_0 \, G} \, m}$ -- taking, for example, the proton leads to $\beta \sim 10^{18}$, and that will give near-maximal values for the special relativistic case.

The spherical shell represents a pretty minimal neutralization model -- we expect more than just falling straight in from our neutralizing particles.  In all of these settings, a more realistic ``cloud" of particles with varying initial velocities and positions would lead to a more realistic (and necessarily longer) neutralization time, and these can be computed numerically in all three of the cases studied here.  The inclusion of a self-force and calculation of a consistent spacetime for astrophysical black holes requires full GR and is an involved task -- our goal here, motivated by the astrophysical observation that macroscopic bodies are neutral, is to provide a simple model that can be studied relatively completely in the classical, special and general relativistic frameworks.

\begin{acknowledgements}
We thank David Griffiths for his early comments on this work -- our anonymous reviewer also provided helpful suggestions.
\end{acknowledgements}


\begin{thebibliography}{18}
\bibitem{note}{It ends up being easier to work directly from the second order equations of motion, our numerical work proceeds from the ODE derived by taking a $\sigma$ derivative of~\refeq{rho2f}.}
\bibitem{GRnote2}{
The event horizon for these charged black holes is not just the usual Schwarzschild one -- the Reissner-Nordstrom horizon is inside the Schwarzschild horizon, so we have not allowed particles to neutralize at the most natural central body radius (using the correct horizon will introduce additional time for the neutralization process).  The use of $r_0 = \frac{2 \, G \, M}{c^2}$ as the radius of our central body just means that in the charged case, we do not have a black hole (there is mass and charge outside of the horizon).
}
\end{thebibliography}
\end{document}